\documentclass[12pt
]{article}

\usepackage{graphicx}
\usepackage{amsfonts}
\usepackage{amsmath}
\usepackage[usenames]{color}
\usepackage{amssymb}
\usepackage[mathscr]{eucal}

\def\Z{\mathbb{Z}}

\def\R{\mathbb{R}}

\begin{document}
\baselineskip 0.6cm
\newcommand{\gsim}{ \mathop{}_{\textstyle \sim}^{\textstyle >} }
\newcommand{\lsim}{ \mathop{}_{\textstyle \sim}^{\textstyle <} }
\newcommand{\vev}[1]{ \left\langle {#1} \right\rangle }
\newcommand{\bra}[1]{ \langle {#1} | }
\newcommand{\ket}[1]{ | {#1} \rangle }
\newcommand{\Dsl}{\mbox{\ooalign{\hfil/\hfil\crcr$D$}}}
\newcommand{\nequiv}{\mbox{\ooalign{\hfil/\hfil\crcr$\equiv$}}}
\newcommand{\nsupset}{\mbox{\ooalign{\hfil/\hfil\crcr$\supset$}}}
\newcommand{\nni}{\mbox{\ooalign{\hfil/\hfil\crcr$\ni$}}}
\newcommand{\EV}{ {\rm eV} }
\newcommand{\KEV}{ {\rm keV} }
\newcommand{\MEV}{ {\rm MeV} }
\newcommand{\GEV}{ {\rm GeV} }
\newcommand{\TEV}{ {\rm TeV} }

\def\diag{\mathop{\rm diag}\nolimits}
\def\tr{\mathop{\rm tr}}

\def\Spin{\mathop{\rm Spin}}
\def\SO{\mathop{\rm SO}}
\def\O{\mathop{\rm O}}
\def\SU{\mathop{\rm SU}}
\def\U{\mathop{\rm U}}
\def\Sp{\mathop{\rm Sp}}
\def\SL{\mathop{\rm SL}}

\def\change#1#2{{\color{blue}#1}{\color{red} [#2]}\color{black}\hbox{}}


\begin{titlepage}
  
\begin{flushright}
  UT-11-46 \\
  IPMU11-0202
\end{flushright}
  
\vskip 2cm
\begin{center}
  
{\large \bf  A Note on Kahler Potential of Charged Matter in F-theory}
  
\vskip 2cm
  
Teruhiko Kawano$^1$, Yoichi Tsuchiya$^1$ and Taizan Watari$^2$

\vskip 0.4cm
    

{\it
  $^1$Department of Physics, University of Tokyo, Tokyo 113-0033, Japan  
  \\[2mm]
  
 $^2$Institute for the Physics and Mathematics of the Universe, 
 University of Tokyo, Kashiwa-no-ha 5-1-5, 277-8583, Japan
  }

\vskip 7cm
\abstract{
We study the Kahler potential of charged matter fields, 
whose profiles have a peak on their matter curve --- 
on an ``intersection'' of 7-branes, in an F-theory compactification. 
It is shown that the Kahler potential is exactly given by the integral 
over the matter curve, but not by the integral over the whole GUT surface 
of 7-branes. 
} 
  
\end{center}
\end{titlepage}



\section{Introduction}

Kahler potential of visible sector particles in supersymmetric theories
has important consequences in physics. Planck-scale suppressed couplings 
between the Standard-Model particles and moduli fields or inflaton 
(may) play the essential role in gravity/anomaly mediated supersymmetry 
breaking and post-inflationary dynamics of the universe. 
However, very few claims can be made about such couplings 
within effective 
field theories below the Planck scale, and string theory is virtually 
the only theoretical framework in the market which enables us to 
make some theoretical progress. 

There are already studies along this line \cite{Kahler}. 
Given the nature of this problem, it is important that string 
compactification with as many realistic features as possible 
should be used for such a study. 
Especially in the context of gravity mediated supersymmetry breaking, 
it is crucial that we use a framework where flavor violation in Yukawa 
couplings has already been understood. 
For this reason, we choose F-theory compactification  
with charged matter fields of the supersymmetric Standard Model
arising from matter curves.
After a few years of intense study on flavor structure of Yukawa matrix, 
it is now known that realistic flavor pattern can be realized in such a
compactification, either by 
tuning one (or a few) complex structure parameter(s) of the matter curve
associated with ${\rm SU}(5)_{\rm GUT}$-{\bf 10} representation 
fields \cite{Hayashi-Flavor}, or by taking appropriate factorization 
limit of geometry\footnote{This is to 
avoid \cite{Caltech-0904, Hayashi-Flavor, Cordova} multiple points of enhanced 
singularity of $E_6$ type and $D_6$ type in the ${\rm SU}(5)_{\rm GUT}$ 
GUT divisor contributing to the low-energy up-type and down-type Yukawa 
matrices. 
It should be noted that factorization of {\it spectral surface} is not enough 
because of the problem discussed in \cite{Tsuchiya, Hayashi-U1,
Grimm-Weigand-U1, Caltech-U1}. The ``factorization'' condition needs to be
implemented as a constraint on global geometry, to say the least. } 
in order to exploit the idea of \cite{HV-Nov08-rev, Harvard-CS}. 
See \cite{Hayashi-Flavor, Hayashi-topology} for yet another solution 
using discrete symmetry. 

In this article, we will make a small progress in this program. 
In F-theory, charged matter fields on ``7-brane intersection'' are 
characterized by holomorphic sections of appropriate line bundles on 
(covering) matter curves \cite{Curio, DI, Penn5, BMRW, Hayashi-1}.
Without a microscopic formulation of F-theory, however, field-theory 
local models on 7+1 dimensions \cite{KV, DW-1, BHV-1, Hayashi-2}
are the only theoretical tool to calculate low-energy effective 
Kahler potential (as well as superpotential). Although the relation 
between the holomorphic sections on matter curves and wavefunctions on 
GUT divisor is now understood, it would be nice if there are expressions 
for observables (like F-term and D-term couplings) that are given directly 
in terms of single-component holomorphic sections on curves, rather than 
those given indirectly through the (possibly multi-component) 
wavefunctions on a surface.
References \cite{Harvard-CS, T-brane} exploited holomorphicity 
in the description of charged matter (and residue integral) and 
independence of F-term couplings on volume moduli, in order to show that 
the F-term Yukawa couplings are given directly by the holomorphic 
sections on the matter curves.  
In this article, we show that the residue integral of 
\cite{Wijn, Harvard-CS, T-brane} can be used also to show 
that the Kahler potential of charged matter fields on matter curves 
is also localized on the matter curves.\footnote{
Reference \cite{BHV-1} wrote down Hermitian inner product of global 
holomorphic sections of line bundles on matter curves as a ``natural'' 
expression, which also follows from simple calculation in the case 
of single-component matter wavefunction with a Gaussian profile in the 
direction transverse to the matter curve. 
Additional evidence in support of the Hermitian inner product on matter
curves was obtained \cite{Hayashi-Flavor} at a special slice of matter
curves where spectral surface is ramified (7-brane monodromy is
non-trivial), and a single-component wavefunction cannot be used 
under any approximation schemes.
This article provides a proof of localization to matter curves  
along the entire matter curves for any ramified spectral surfaces.}

\section{Problem}

Local geometry of Calabi--Yau 4-fold $X_4$ for supersymmetric 
compactification of F-theory is translated into field theory 
model on a patch of 7+1-dimensional spacetime \cite{KV, DW-1, BHV-1}.
When an elliptic fibration $\pi_X: X_4 \rightarrow B_3$ with a section 
$\sigma: B_3 \rightarrow X_4$ has a locus of singularity of type 
$G''$ (such as $G'' = A_4$) in the transverse direction at a generic 
point of a complex surface 
$S_{\rm GUT} \hookrightarrow B_3 \hookrightarrow X_4$ appearing as an 
irreducible component of the discriminant of the elliptic fibration, 
the field theory model for a local geometry in $X_4$ containing 
an open local patch $U \subset S_{\rm GUT} \subset X_4$ is a supersymmetric 
gauge theory on $\R^{3,1} \times U$ with the gauge group $G$ containing 
$G^{''}$. The choice of the gauge group $G$ depends on the local
geometry, as well as on the level of approximation we want in the 
field theory model. 

In a field theory local model with the gauge group $G = E_6$ that 
is broken down to $G'' = {\rm SU}(5)_{\rm GUT}$, for example, 
the symmetry breaking 
(deformation of singularity) is encoded in a non-trivial configuration 
of gauge field $A_m$ and Higgs field $\varphi$ in the 
$G_{\rm str} = {\rm U}(2)$ subgroup (subalgebra) in $E_6$ commuting 
${\rm SU}(5)_{\rm GUT}$. 
Chiral multiplets in the $R_I={\bf 10}$ and $R_I=\bar{\bf 5}$
representations of $G^{''}={\rm SU}(5)_{\rm GUT}$ describing low-energy
effective theory below the Kaluza--Klein scale have, in the field theory
local model on $\R^{3,1} \times U$, their wavefunction profiles 
$(\delta A, \delta \varphi)$ in the corresponding irreducible components 
\begin{eqnarray}
 {\rm Res}^{G = E_6}_{\U(2) \times {\rm SU}(5)_{\rm GUT}} \mathfrak{g} & = &
 ({\bf 2}\otimes \bar{\bf 2}, {\bf 1}) + ({\bf 1}, {\bf adj.}) + 
 \left[ ({\bf 2}, {\bf 10}) + (\wedge^2 {\bf 2}, \bar{\bf 5})\right] +
 {\rm h.c.}, \\
 & = & ({\bf 2} \otimes \bar{\bf 2}, {\bf 1})+({\bf 1}, {\bf adj.}) 
 + \oplus_I (U_I, R_I).
\end{eqnarray}
The zero mode wavefunctions 
$(\delta A^{(0,1)}, \delta \varphi^{(2,0)}) \equiv (\psi, \chi)$ 
in the irreducible component $(U_I, R_I)$ should satisfy 
\begin{eqnarray}
 \omega \wedge D' \psi + 
 \frac{|\alpha|^2}{2} \rho_{U_I}(\vev{\overline{\varphi}}) \chi & = & 0, 
     \label{eq:0-eq-D}\\
 D'' \psi & = & 0,  \label{eq:0-eq-H} \\
 D'' \chi - i \rho_{U_I}(\vev{\varphi}) \psi & = & 0. \label{eq:0-eq-G}
\end{eqnarray}
Here, $\omega$ is a Kahler form on $U \subset S_{\rm GUT}$, and 
$D'$ and $D''$ are the $(1,0)$ and $(0,1)$ part of covariant derivative 
for fields in the representation $U_I$ of $G_{\rm str}$. 
See \cite{Hayashi-Flavor} (and \cite{Hayashi-2}) for notations and 
conventions adopted in this article. 

Chiral multiplets in a representation $R_I$ in low-energy
effective theory can be characterized only in a geometry along a matter 
curve $\bar{c}_{(R_I)} \hookrightarrow S_{\rm GUT}$ corresponding to the
representation $R_I$. They form a vector 
space corresponding to global holomorphic sections of certain line
bundles on covering matter curve $\tilde{\bar{c}}_{(R_I)}$ 
(see \cite{Hayashi-1} for the difference between matter curves 
$\bar{c}_{(R_I)}$ and their corresponding covering matter curves 
$\tilde{\bar{c}}_{(R_I)}$; we will not focus on the difference in this
article, however.):  
\begin{equation}
 H^0 \left( \tilde{\bar{c}}_{(R_I)}; {\cal L}_{G^{(4)}} \otimes 
   K^{1/2}_{\tilde{\bar{c}}_{(R_I)} } \right).
\label{eq:matter-linebundle-on-curve}
\end{equation}
Low-energy effective field theory is described by choosing a
basis $\{ \tilde{f}_{(R_I); i} \}$ of this vector space; 
the choice of basis of this vector space sets a basis 
in the chiral multiplets $\{\Phi_{(R_I); i} \}$ of the low-energy
physics. $\{ \Phi_{(R_I); i}\}$'s (and $\{\tilde{f}_{(R_I);i} \}$'s)
have their corresponding 
wavefunctions $(\psi_{i}, \chi_i)$ for any local patch 
$U \subset S_{\rm GUT}$ containing a segment of $\bar{c}_{(R_I)}$. 
The kinetic terms of the chiral multiplets in the representation $R_I$
are given in terms of these wavefunctions as in \cite{Hayashi-Flavor} by 
\begin{eqnarray}
 K_{\rm eff.} & = & K^{(R_I)}_{i\bar{j}} \; 
 \Phi^{\dagger}_{(R_I); \bar{j}} e^{\rho_{R_I}(V)} \Phi_{(R_I); i} + \cdots, \\ 
 K_{i\bar{j}}^{(R_I)} & \simeq & \frac{c_{(U_I,R_I)} M_*^4}{4\pi} \int_U 
i \omega \wedge \left[ \bar{\psi}_{\bar{j}} \wedge \psi_i \right]
+ \frac{|\alpha|^2}{2} \left[ \overline{\chi}_{\bar{j}}\wedge \chi_i \right].
\label{eq:K.F.-truncation}
\end{eqnarray}
Here, in (\ref{eq:K.F.-truncation}), the zero-mode wavefunctions 
$(\psi_i,\chi_i)$ are treated as 
$(r_I \equiv {\rm dim}. \; U_I)$-component fields, and 
$r_I \times r_I$ unit matrix is used for the inner product. 
The couplings between the visible sector particles such as quarks and
leptons and various moduli fields (including Kahler moduli, complex 
structure moduli, right-handed neutrinos) of the Calabi--Yau 4-fold 
are encoded in the moduli value dependence of this kinetic mixing 
matrix\footnote{Here, the ellipses in the first line are the terms 
in higher order in the charged chiral multiplets $\Phi_i$. We should note that 
the equations (\ref{eq:0-eq-D}--\ref{eq:0-eq-G}) implicitly assume 
that the zero-mode wavefunctions are infinitesimal fluctuations from 
the background in $\mathfrak{g}_{\rm str}$ rather than finite deformations.  
The expression (\ref{eq:K.F.-truncation}) for the kinetic mixing matrix 
$K^{(R_I)}_{i\bar{j}}$ corresponds to a partial contribution from 
an open patch $U \subset S_{\rm GUT}$, and all such contributions 
from patches along the matter curve $\bar{c}_{(R_I)}$ should be 
summed up. We should also note that this expression only takes account 
of the leading order term in the 7+1-dimensional field theory local
model Lagrangian in the $1/[M_*^4 \times {\rm vol}(S_{\rm GUT})]$ expansion. 
It should also be remembered that the low-energy effective Kahler
potential should not be obtained by just truncating (forgetting about) 
various Kaluza--Klein modes, but by integrating them out. This
difference should matter for the Kahler potential. 
In this article, as well as in \cite{Hayashi-Flavor}, however, we will 
take a naive dimensional reduction as a first step to see what 
is going on.} $K^{(R_I)}_{i\bar{j}}$.

The problem we face in this article is the following. 
The expression (\ref{eq:K.F.-truncation}) is given by an integration 
over a complex {\em surface} $S_{\rm GUT}$, not precisely on the 
matter {\em curve} $\tilde{\bar{c}}_{(R_I)}$, although the wavefunctions 
$(\psi_i, \chi_i)$ become small quickly at points far away from the
matter curve. We show by using residue integral 
(as in \cite{Wijn, Harvard-CS, T-brane}) that there is an expression 
for the kinetic mixing matrix where the integration is carried out 
exactly on the matter curve. 

\section{Use of Residue Integral}

The zero-mode wavefunctions $(\psi, \chi)$ on a complex surface $U$ 
are related to the sections $\tilde{f}$ on complex curve 
$\tilde{\bar{c}}_{(R_I)}$ as follows \cite{Hayashi-2, DW-3, Harvard-CS, 
Hayashi-Flavor}. Hereafter, we drop reference to representations such as 
$U_I, R_I$ and $\rho_{U_I}$ from zero-mode wavefunctions, sections and 
background field configuration; the following argument is applied to 
any one of irreducible components $(U_I, R_I)$, and should be applied 
separately to these components one by one. 
Now, first, because the zero-mode wavefunctions $(\psi,\chi)$ are 
vector bundle $V_{U_I}$ (${\rm rank} \; V_{U_I} = {\rm dim} \; U_I = r_I$) 
valued $(0,1)$-form and $(2,0)$-form, respectively, 
we can take arbitrary frame for local
trivialization of the vector bundle $V_{U_I}$ in describing them.  
When the unitary frame is replaced by a holomorphic frame, 
\begin{equation}
 \psi = {\cal E} \tilde{\psi}, \qquad \chi = {\cal E} \tilde{\chi},
\label{eq:cpxfd-gauge-transf}
\end{equation}
where ${\cal E}$ is an $r_I \times r_I$ matrix whose 
value is in complexification of the structure group $G_{\rm str.}$, 
and the zero-mode equations (\ref{eq:0-eq-D}--\ref{eq:0-eq-G}) 
become\footnote{The background field configuration $\vev{\varphi}$ is 
now simply denoted by $\varphi$, as the distinction between zero-mode 
wavefunctions $(\psi, \chi)$ and background configuration in a
representation $U_I$, $(A, \varphi)$, will already be clear.}
\begin{eqnarray}
 \omega \wedge \tilde{D}' \tilde{\psi} + \frac{|\alpha|^2}{2} 
   \widetilde{\bar{\varphi}} \; \tilde{\chi} 
 & = & 0, \label{eq:0-eq-D-hol}\\
 \bar{\partial} \widetilde{\psi} & = & 0, \label{eq:0-eq-H-hol} \\
 \bar{\partial} \tilde{\chi}
 - i \widetilde{\varphi} \; \tilde{\psi}
  & = & 0. \label{eq:0-eq-G-hol}
\end{eqnarray}
Hermitian inner product of sections of $V_I$ is given in this
holomorphic frame by an $r_I \times r_I$ matrix 
$H = {\cal E}^\dagger {\cal E}$.

The F-term conditions (\ref{eq:0-eq-H-hol}, \ref{eq:0-eq-G-hol}) 
imply that the zero-mode wavefunctions are given by some 
$V_{U_I}$-valued field\footnote{This $\tilde{\Lambda}$ wavefunction
multiplied by a 4D fluctuation $\phi(x)$ takes its value 
in the $(U_I, R_I)$ component of $\mathfrak{g}$, and can also be
regarded as the generator of an infinitesimal complexified gauge transformation
beyond ${\cal E}$ in (\ref{eq:cpxfd-gauge-transf}).} 
$\tilde{\Lambda}$ and $\tilde{f} \in \Gamma(U; V_{U_I} \otimes K_U)$:
\begin{equation}
 \tilde{\psi}  =  \bar{\partial} \; \tilde{\Lambda}, \qquad 
   \widetilde{\chi} = i \widetilde{\varphi} \; 
 \tilde{\Lambda} + \tilde{f}_{},
\end{equation}
with a redundancy of the form 
$\tilde{\Lambda} \rightarrow \tilde{\Lambda} - k$ and 
$\tilde{f} \rightarrow \tilde{f} + i \widetilde{\varphi} \; k$ for 
arbitrary $k \in \Gamma(U; V_{U_I})$.
The remaining D-term condition (\ref{eq:0-eq-D-hol}) sets the relation 
between $\tilde{\Lambda}$ and holomorphic $\tilde{f}$:
\begin{equation}
 \omega \wedge H^{-1} \partial 
     \left( H \left( \bar{\partial} \tilde{\Lambda}\right) \right)
 + \frac{|\alpha|^2}{2} H^{-1} \widetilde{\varphi}^\dagger H 
     \left(i \widetilde{\varphi} \; \tilde{\Lambda} + \tilde{f}
     \right) = 0. \label{eq:0-eq-D-fLambda}
\end{equation}
Thus, the zero-mode wavefunctions are determined from 
$\tilde{f} \in \Gamma(U; V_{U_I} \otimes K_U)$, and the redundancy 
we mentioned earlier allows us to identify the vector space of zero
modes (\ref{eq:matter-linebundle-on-curve}) characterized on matter 
curves with \cite{Hayashi-2, DW-3, Harvard-CS, Hayashi-Flavor}
\begin{equation}
{\rm Coker}\left[(\varphi \times) : 
  H^0(\widetilde{C}_{U_I}; \widetilde{\cal N}_{U_I}) \rightarrow 
  H^0(\widetilde{C}_{U_I}; \widetilde{\cal N}_{U_I} \otimes 
    \tilde{\pi}^*_{C_{U_{I}}}(K_{U})) \right].
\label{eq:matter-coker}
\end{equation}
$\widetilde{C}_{U_I}$ is the (desingularization of the) spectral surface 
for the $K_U$-valued Higgs bundle of the vacuum configuration 
in the $G_{\rm str}$-$U_I$ representation, and 
$\widetilde{\cal N}_{U_I}$ is a line bundle on it, keeping information 
of 4-form flux in F-theory compactification. Through the projection 
$\tilde{\pi}_{C_{U_I}} : \widetilde{C}_{U_I} \rightarrow U$, the vector
bundle $V_{U_I}$ and the line bundle $\widetilde{\cal N}_{U_I}$ are
related as follows: $V_{U_I} = \tilde{\pi}_{C_{U_I} *}
\left(\widetilde{\cal N}_{U_I} \right)$. 
See \cite{Hayashi-2} for more information.

Having prepared all the necessary material, let us now rewrite the 
kinetic mixing matrix as follows.\footnote{In order to save space,
we will drop the representation-dependent and calculable constant 
$c_{(U_I, R_I)}$ in (\ref{eq:K.F.-truncation}) in the rest of this
article. It is dimensionless, and is of order unity.}
\begin{eqnarray}
 K_{i\bar{j}}^{(R_I)} & \simeq & \frac{  M_*^4}{4\pi} \int_U 
  i \omega \wedge \left[ (\partial \tilde{\Lambda}^\dagger_{\bar{j}})
		   \wedge H (\bar{\partial} \tilde{\Lambda}_i) \right] + 
  \frac{|\alpha|^2}{2}\left[ 
   \left\{ \tilde{f}^\dagger_{\bar{j}}
    -i\tilde{\Lambda}^\dagger_{\bar{j}}
    \left(\widetilde{\varphi}\right)^\dagger  \right\} \wedge H 
   \left\{ \tilde{f}_i + i \widetilde{\varphi} \tilde{\Lambda}_i \right\}
  \right] \nonumber \\
 & = & \frac{ M_*^4}{4\pi} \int_U
    \partial \left( i \omega \wedge 
      \left[ \tilde{\Lambda}^\dagger_{\bar{j}} \wedge
                H (\bar{\partial} \tilde{\Lambda}_i )
      \right] \right)
   +   \frac{|\alpha|^2}{2}\left[ 
    \tilde{f}^\dagger_{\bar{j}}  \wedge H 
   \left\{ \tilde{f}_i + i \widetilde{\varphi} \tilde{\Lambda}_i \right\}
  \right] \nonumber \\
 & + & \frac{ M_*^4}{4\pi} \int_U -i \omega \wedge
   \left[ \tilde{\Lambda}^\dagger_{\bar{j}} \wedge
     \partial \left(H  (\bar{\partial} \tilde{\Lambda}_i)\right) \right]
 -i \frac{|\alpha|^2}{2}\left[ \tilde{\Lambda}^\dagger_{\bar{j}}
     (\widetilde{\varphi})^\dagger \wedge H 
     \left\{ \tilde{f}_i + i \widetilde{\varphi} \tilde{\Lambda}_i \right\}
     \right]. 
\end{eqnarray}
The last line vanishes because of the D-term 
condition (\ref{eq:0-eq-D-fLambda}), and the first term can be dropped 
because it is a total derivative. 
Let us now take a set of local coordinates $(u,v)$ on $U \subset S_{\rm GUT}$, 
and describe 
\begin{equation}
 \tilde{f} = \tilde{f}_{uv} du \wedge dv, \qquad 
 \widetilde{\varphi} = \widetilde{\varphi}_{uv} du \wedge dv.
\end{equation}
Inserting an $r_I \times r_I$ matrix 
$(\widetilde{\varphi}_{uv})^{\dagger -1} (\widetilde{\varphi}_{uv})^\dagger$, 
we find that 
\begin{eqnarray}
 K^{(R_I)}_{i\bar{j}} & \simeq & \frac{ M_*^4}{4\pi} 
   \int_U   \frac{|\alpha|^2}{2}\left[ 
    \tilde{f}^\dagger_{\bar{j}}  \wedge H 
   \left\{ \tilde{f}_i + i \widetilde{\varphi} \tilde{\Lambda}_i \right\}
  \right], \nonumber \\
 & = &  \frac{ M_*^4}{4\pi} \int_U  \frac{|\alpha|^2}{2}\left[ 
    ((\tilde{f}_{j})_{uv})^\dagger 
    (\widetilde{\varphi}_{uv})^{\dagger -1} (\widetilde{\varphi})^\dagger
 \wedge H 
   \left\{ \tilde{f}_i + i \widetilde{\varphi} \tilde{\Lambda}_i \right\}
  \right], \label{eq:step-A} \\
 & = & - \frac{ M_*^4}{4\pi} \int_U \left[ 
        ((\tilde{f}_{j})_{uv})^\dagger 
        (\widetilde{\varphi}_{uv})^{\dagger -1}
        \omega \wedge \partial \left(H \tilde{\psi}_i \right)
\right], \label{eq:step-B} \\
 & = & \frac{ M_*^4}{4\pi} \int_U \left[
    \partial \left\{ ((\tilde{f}_{j})_{uv})^\dagger 
        (\widetilde{\varphi}_{uv})^{\dagger -1} \right\} 
    \omega \wedge H \tilde{\psi}_i 
          \right]; \label{eq:step-C}
\end{eqnarray}
the D-term condition (\ref{eq:0-eq-D-fLambda}) was used in between 
(\ref{eq:step-A}) and (\ref{eq:step-B}) once more. 

It is now easy to see that this expression for the kinetic mixing 
matrix is localized precisely on the matter curve. Because  
$\tilde{f}$ and $\widetilde{\varphi}$ are holomorphic sections of 
$K_{U} \otimes V_{U_I}$ and $K_U \otimes {\rm End} V_{U_I}$,
respectively, the $r_I$-component field 
$[(\widetilde{\varphi}_{uv})^{-1} \cdot \tilde{f}_{uv}]$ remains
holomorphic everywhere in $U \subset S_{\rm GUT}$, except where 
${\rm det} (\widetilde{\varphi}_{uv})$ vanishes and the inverse matrix 
is not well-defined, namely, the matter curve for the $(U_I, R_I)$ component.
If the matter curve is the simple zero of the determinant, then 
$\bar{\partial}$ of the $r_I$-component field 
$[(\widetilde{\varphi}_{uv})^{-1} \cdot \tilde{f}_{uv}]$ becomes a 
$(0,1)$-form with its support precisely on the matter curve. 
This procedure is quite similar to the use of residue integral 
in \cite{Wijn, Harvard-CS, T-brane}  for the proof of localization of 
F-term Yukawa
couplings on the matter curve. Although we cannot always expect that
holomorphicity plays some role in couplings in the Kahler potential, 
in this case, we can still use the holomorphicity associated 
with (\ref{eq:matter-coker}) and residue integral to show the precise
localization on the matter curve. 

\subsection{Residue Integral for Rank-2 Ramified Spectral Surface}

To see the power of the expression (\ref{eq:step-C}), 
let us see more in detail how (\ref{eq:step-C}) is like when 
the field theory local model is given by a rank-2 ramified spectral
cover $C_{U_I}$:
\begin{equation}
 (\xi_{uv})^2 + 2 c'M_*  (M_*u) (\xi_{uv}) - c^2 M_*^2 (M_* v) = 0, 
\label{eq:spec-surf}
\end{equation}
where $(u,v)$ are local coordinates on $U \subset S_{\rm GUT}$, 
and $\xi_{uv}$ is the fiber coordinate of the canonical bundle $K_U$ with the
trivialization frame $du \wedge dv$. $c$ and $c'$ are dimensionless
constants, which carry the complex structure information of the
Calabi--Yau 4-fold of F-theory compactification. 
In fact, this example is practically important, because 
the zero-mode wavefunctions of chiral multiplets in the $R_I={\bf 10}$ 
representation [$R_I = \bar{\bf 5}$ or ${\bf 5}$ representation] of 
$G^{''}={\rm SU}(5)_{\rm GUT}$ are determined by the Higgs field
background described by (\ref{eq:spec-surf}) near the points of 
enhanced singularity of type $E_6$ [resp. type $A_6$].

Holomorphic sections of {\it line} bundles 
$\tilde{f} \in \Gamma(C_{U_I}; {\cal N}_{U_I} \otimes \pi_{C_{U_I}}^* (K_U))$
can be translated into the ${\rm dim} \; U_I = 2$ component 
$(2,0)$-forms on $U \subset S_{\rm GUT}$ as follows. First, we take a
holomorphic frame $e_N$ for local trivialization of ${\cal N}_{U_I}$, 
so that holomorphic sections $\tilde{f}$ can be written in terms of 
a holomorphic function $\tilde{f}_{uv}(\xi_{uv},u)$:
\begin{equation}
 \tilde{f} = \tilde{f}(\xi_{uv},u) \; 
  e_N \otimes \pi_{C_{U_I}}^* (du \wedge dv).
\end{equation}
A $\Z_2$-Galois transformation 
$[\xi_{uv} + c' M_*(M_* u)] \rightarrow - [\xi_{uv} + c' M_*(M_* u)]$ 
of the quadratic equation acts on the 2-fold cover (\ref{eq:spec-surf}), 
and holomorphic functions $\tilde{f}(\xi_{uv},u)$ on the spectral cover 
can be written as a sum of $\Z_2$-even part and odd part: 
\begin{equation}
 \tilde{f}_{uv}(\xi_{uv},u) =
   \tilde{f}^{(+)}_{uv}(v,u)
 + M_*^{-1} (\xi_{uv} + c' M_* (M_* u)) \tilde{f}^{(-)}_{uv}(v,u). 
\end{equation}
A 2-component description of the holomorphic sections 
of $\pi_{C_{U_I}*} \left({\cal N}_{U_I} \right)\otimes K_U 
 = V_{U_I}\otimes K_U$ is provided by 
\begin{equation}
 \tilde{f}_{uv} = \left( \tilde{f}^{(+)}_{uv}, 
                       \; \tilde{f}^{(-)}_{uv} \right)^T.
\end{equation}
When the 2-component description of 
$V_I = \pi_{C_{U_I}*} \left({\cal N}_{U_I} \right)$ 
is also given by using the Galois $\Z_2$-even and odd components, 
the multiplication of $\xi_{uv} = 2\alpha \widetilde{\varphi}_{uv}$ 
in (\ref{eq:matter-coker}) is represented by 
\cite{Hayashi-Flavor}\footnote{Such a configuration of
spectral cover was named ``T-brane'' in \cite{T-brane}.}
\begin{equation}
 2\alpha \varphi_{uv} = M_* \left( \begin{array}{cc}
	 -c' (M_*u) & c^2 (M_* v) + (c')^2 (M_* u)^2 \\
         1 & -c' (M_*u)
			       \end{array}\right).
\end{equation}

Now it is straightforward to calculate 
\begin{equation}
 \left(\widetilde{\varphi}_{uv}\right)^{-1} \tilde{f}_{uv}
  = \frac{2\alpha}{M_* c^2 (M_* v)}
  \left( \begin{array}{cc}
	 c' (M_*u) & c^2 (M_* v) + (c')^2 (M_* u)^2 \\
         1 & c' (M_*u)
			       \end{array}\right)
  \left( \begin{array}{c}
   \tilde{f}^{(+)}_{uv}(v,u) \\ \tilde{f}^{(-)}_{uv}(v,u) 
	 \end{array}\right),
\end{equation}
and hence 
\begin{equation}
 \bar{\partial} \left\{  \left(\widetilde{\varphi}_{uv}\right)^{-1}
		 \tilde{f}_{uv} \right\} = 
  \frac{2\alpha}{M_*^{2} c^2} \pi \delta^2(v,\bar{v}) d\bar{v} 
   \left( \begin{array}{c}
    c' (M_* u) \\ 1
	  \end{array}\right) 
   [\tilde{f}_{uv}(\xi_{uv},u)]|_{\xi_{uv} = 0}.
\label{eq:pick-up-residue}
\end{equation}
This expression is used in (\ref{eq:step-C}). This is how we can see 
that the kinetic mixing matrix is reduced to an expression along 
the matter curve $v = 0, {}^\forall u$ in $U \subset S_{\rm GUT}$.
It is also nice to see that the single component wavefunction on the 
curve, $[\tilde{f}_{uv}(\xi_{uv},u)]_{\xi=0}$ to be identified directly with 
an element of (\ref{eq:matter-linebundle-on-curve},
\ref{eq:matter-coker}), can be pulled out from the $2 \times 2$ 
vector--matrix calculation. 

\subsection{Expressions at the $E_6$ and $A_6$ points:
--use of Painleve equation--}

Although it is an important step to show that the 
expression (\ref{eq:K.F.-truncation}) is reduced to an integral
precisely on matter curves, there still remains another important task 
of rewriting the integrand (\ref{eq:step-C}) on the curve in terms of 
$\tilde{f}_{uv}$ and Kahler and complex structure moduli of the 
Calabi--Yau 4-fold. 

There are not many field-theory local models, however, where 
the relation between $\tilde{\psi}_i$ and $\tilde{f}_i$ and moduli dependence 
of the Hermitian inner product $H$ is understood very well. 
In a field-theory local model where a rank-1 extension of 
the gauge group provides a good approximation of the 4-fold 
geometry \cite{KV, DW-1, BHV-1}, zero-mode wavefunctions $(\psi,\chi)$ 
have Gaussian profile 
in the direction transverse to the matter curve, and integration 
of (\ref{eq:K.F.-truncation}) can be carried out in the transverse
direction, first. Such a result---partial contribution 
to $K^{(R_I)}_{i\bar{j}}$ from segments of matter curves satisfying the
condition above---is \cite{Hayashi-Flavor}
\begin{equation}
 \Delta K^{(R_I)}_{i\bar{j}} \propto \omega_{\bar{c}_{(R_I)}} \;
 \left[
   (2\alpha (\tilde{f}_j)_{uv})^* 
   \frac{\sqrt{h^{\bar{u}u}} M_*^4}{2|F|}
   (2\alpha \tilde{f}_i)_{uv} 
 \right], 
\label{eq:result-Gauss}
\end{equation} 
where we assume that the spectral cover of the field-theory local model 
is given by $\xi_{uv} - F v = 0$.  
This result can be reproduced easily also from (\ref{eq:step-C}) by
using the relation between $\tilde{\psi}$ and $\tilde{f}_{uv}$ in the 
Gaussian zero modes. 

There is yet another case where we know about the background Higgs bundle 
configuration and zero-mode wavefunction profiles well \cite{Hayashi-Flavor, T-brane}. 
That is the $u=0$ slice of the field-theory local model given 
by (\ref{eq:spec-surf}), and will be considered in this subsection. 
It also describes the configuration of the field-theory local model 
near the $E_6$ and $A_6$ points, the former of which gives rise to the 
up-type Yukawa couplings. In the rest of this section, we will 
extend the analysis of \cite{Hayashi-Flavor, T-brane} 
by using the residue integral (\ref{eq:step-C}) 
to obtain an easy-to-use expression for the integrand. 
We assume that the Kahler metric $\omega$ is diagonal in the
$(u,v)$ coordinates in $U \subset S_{\rm GUT}$, that is, 
$h_{u\bar{v}} = h_{v\bar{u}} = 0$, and the diagonal terms 
$h_{u\bar{u}}$ and $h_{v\bar{v}}$ remain constant.

In this special case, at the $u=0$ slice of the local patch $U$,  
a special ansatz can be used for the frame-change $2 \times 2$ matrix
${\cal E}$ (complexified gauge transformation) 
in (\ref{eq:cpxfd-gauge-transf}):
\begin{equation}
 {\cal E} = \left(\begin{array}{cc}
	     \underline{\cal E}^{-1} & \\
             & \underline{\cal E} \; e_{uv} 
		  \end{array}\right).
\end{equation}
Here, this ansatz was generalized slightly from 
\cite{Hitchin-1, Hayashi-Flavor, T-brane} by introducing 
a factor $e_{uv}$ in order to obtain 
an expression that is covariant under rescaling of local coordinates, 
$(u',v') = (\lambda_u u, \lambda_v v)$; 
the factor $e_{uv}$ needs to scale as 
$e_{uv}' = e_{uv} / (\lambda_u \lambda_v)$.
Hitchin equation (D-term BPS condition) turns into 
\begin{equation}
 \partial_v \bar{\partial}_{\bar{v}}
      \ln \left(|e_{uv}|^2 |c|^{-2} {\cal H}^2 \right) = 
 |\underline{c}|^2 M_*^2
  \left[ \left(|e_{uv}|^2 |c|^{-2} {\cal H}^2 \right) - 
         \frac{|M_* v|^2}{|e_{uv}|^2 |c|^{-2} {\cal H}^2}
  \right]  
\label{eq:BPS-D}
\end{equation}
under this ansatz, where 
${\cal H} \equiv \underline{\cal E}^\dagger \underline{\cal E}$
and $\underline{c} \equiv c /\sqrt{h_{u\bar{u}}}$.
Requiring that ${\cal H}/|v|^{1/2}$ be bounded at large $|v|$ and smooth 
at $v = 0$ for physical reasons, one can see that this ${\cal H}$ 
(and hence the $2 \times 2$ Hermitian inner product) is given by 
a single function ${\cal H}_*$ that depends only on a combination 
$t'\equiv |M_* v|^{3/2} |\underline{c}|$ \cite{Hayashi-Flavor}.
\begin{equation}
|e_{uv}|^2 |c|^{-2} {\cal H}^2 = |\underline{c}|^{-2/3} {\cal H}_*(t').
\end{equation}
Reference \cite{T-brane} pointed out that the differential equation 
(\ref{eq:BPS-D}) is seen as the Painleve III equation, where 
$|e_{uv}| |c|^{-1} {\cal H} / |M_* v|^{1/2}$ is the unknown function 
and $t = (8/3) t'$ is the variable.
From the literature on Painleve equation \cite{Painleve-III,
Zamolodchikov}, the numerical value of 
$H_{0*} \equiv {\cal H}_*^2(|v|=0) \simeq 0.53$ for the bounded solution
in \cite{Hayashi-Flavor} turns out to be 
\begin{equation}
 H_{0*} = 3^{2/3} 
  \left(\frac{\Gamma(2/3)}{\Gamma(1/3)} \right)^2 \simeq 0.531457\ldots.
\end{equation}
This determines the Hermitian inner product $H$ in (\ref{eq:step-C}):
\begin{equation}
 H = \left(\begin{array}{cc}
      {\cal H}_*^{-1} \times |\underline{c}|^{1/3}/|c| & \\
  & {\cal H}_* \times |c|/|\underline{c}|^{1/3} 
	   \end{array}\right) |e_{uv}|.
\label{eq:inn-prod-E6}
\end{equation}

The remaining factor in (\ref{eq:step-C}) is the zero-mode wavefunction 
$\tilde{\psi}$ in the lower ($\Z_2$-odd) component. It is related to 
$\tilde{f}_{uv}$ in the $u=0$ slice as \cite{Hayashi-Flavor}
\begin{equation}
 \left. \tilde{\psi}^{(-)}_{\bar{v}}d\bar{v}  \right|_{v=0} =
  i \frac{|\underline{c}|^{4/3} }{c^2}
 \left(- \frac{A_1}{A_0} \right)_* [2\alpha\tilde{f}_{uv}]_{v=0} \;  d\bar{v},
\label{eq:psi-at-E6}
\end{equation}
where $A_{0,1}$ specifies the linear combination of two independent 
modes of the zero-mode equations: 
$\tilde{\chi}^{(+)}(|v|^2) = A_0 + A_1 |M_* v|^2 + \cdots$.
The ratio $(A_1/A_0)_* \equiv (A_1/A_0) \times |\underline{c}|^{-4/3}$ 
should be determined from the requirement 
that the zero-mode wavefunctions contain only decaying mode (without
growing mode) in a region far away from the matter curve.
See the appendix C of \cite{Hayashi-Flavor} for more details. 
Thus, using (\ref{eq:pick-up-residue}, \ref{eq:inn-prod-E6},
\ref{eq:psi-at-E6}) in (\ref{eq:step-C}), we obtain 
\begin{equation}
 \Delta K_{i\bar{j}}^{(R_I)} \simeq \omega_{\bar{c}_{(R_I)}} \; 
  \left[
  (2\alpha (\tilde{f}_j)_{uv})^* 
  \frac{\sqrt{h^{\bar{u}u}} |e_{uv}| M_*^2}{2|c|^2}
 (2\alpha (\tilde{f}_i)_{uv}) \right] \; 
 \sqrt{H_{0*}} \left(- \frac{A_1}{A_0}\right)_*;
\end{equation}
this result is regarded as a small improvement from the one in the appendix C 
of \cite{Hayashi-Flavor}, in that the factor $e_{uv}$ was introduced, 
so that this expression is invariant under the rescaling of local
coordinates $(u',v') = (\lambda_u u, \lambda_v v)$.
Reference \cite{T-brane} pointed out that the damping boundary condition 
of the zero-mode wavefunctions corresponds to 
\begin{equation}
 \left(-\frac{A_1}{A_0}\right)_* = \frac{1}{\sqrt{H_{0*}}}.
\end{equation}
Therefore, an infinitesimal contribution to the kinetic mixing matrix 
of ${\rm SU}(5)_{\rm GUT}$-{\bf 10} representation 
[of ${\rm SU}(5)_{\rm GUT}$-$\bar{\bf 5}$ and ${\bf 5}$ representations] 
from each $E_6$ singularity point [$A_6$ singularity
point\footnote{Study of \cite{Esole-Yau} shows that the singular elliptic
fiber on the ``$A_6$-singularity points'' in the small resolution of
Calabi--Yau 4-fold is not of $I_6$ type. We still maintain the same 
terminology, $A_6$-singularity point, precisely in the meaning 
of \cite{Hayashi-2}. After all, without a microscopic formulation of
F-theory, there is no persuasive argument (rather than a 
try-and-error approach) about how to deal with singular geometry 
{\it for physics purposes}, and how to derive low-energy physics from 
singular or desingularized geometry. In order to bypass this
complication, physics result of Heterotic string theory was translated 
into F-theory through duality, and it was discovered that 
$E_6$ [$A_6$ resp.] gauge theory with a smooth Higgs background 
(without any singularities) reproduces the physics result translated 
from Heterotic string \cite{Hayashi-1, Hayashi-2, Hayashi-Flavor}. 
This is enough in providing theoretical basis for most of the 
applications to low-energy physics, if not all. 
There still remains as a fundamental question, however, what kind of 
geometry ``strings'' or membranes see precisely.} resp.] becomes quite simple:
\begin{equation}
 \Delta K^{(R_I)}_{i\bar{j}} = \omega_{\bar{c}_{(R_I)}} \;
  \left[
  (2\alpha (\tilde{f}_j)_{uv})^* 
  \frac{\sqrt{h^{\bar{u}u}} |e_{uv}| M_*^2}{2|c|^2}
 (2\alpha (\tilde{f}_i)_{uv})
  \right].
\label{eq:result-ramif-pnt}
\end{equation}

\section{Discussions}

In this article, we have shown that the Kahler potential 
of charged matter localized on matter curves can be reduced 
to integrals over the matter curves, but not over the whole 
GUT surface $S_{\rm GUT}$. 
This observation itself is almost obvious in Type IIB Calabi--Yau 
orientifold compactifications, because there is a zero mode 
in the world-sheet field $X^M(\sigma, \tau)$ only in the directions
where Neumann condition is imposed at both $\sigma = 0$ and 
$\sigma = \pi$ boundaries. But in F-theory, where such a world-sheet 
formulation is absent, a proof was necessary for the existence 
of an expression for the kinetic mixing matrix (like (\ref{eq:step-C})) 
that is localized along the matter curves.

In order to study how the charged matter Kahler potential 
(kinetic mixing matrix) depends on Kahler moduli and complex 
structure moduli, however, further work needs to be done. 
Charged matter chiral multiplets are characterized by 
single component holomorphic sections $\tilde{f}$ along 
the matter curves, but detailed and analytic description 
has not been given to the relation between $H \tilde{\psi}$ 
and $\tilde{f}$, and to how the relation depends on the bulk 
Calabi--Yau 4-fold moduli. So far, only partial results, 
(\ref{eq:result-Gauss}) and (\ref{eq:result-ramif-pnt}), 
have been obtained. 

There is yet another open problem associated with charged 
matter kinetic mixing matrix (Kahler potential). 
This problem may be interesting from a theoretical perspective. 
For charged matter fields arising from the bulk of a stack of multiple
7-branes (so that they are in the adjoint representation of the 7-brane 
gauge group), chiral multiplets are characterized either 
as $H^1(S_{\rm GUT}; {\cal O})$ or 
$H^0(S_{\rm GUT}; K_S) = [H^2(S_{\rm GUT}; {\cal O})]^\times$. 
Chiral multiplets of the first kind are described by wavefunctions 
$\psi$, and those of the second kind by wavefunctions $\chi$ 
(e.g. \cite{DW-1, BHV-1} in F-theory compactifications).
As studied in \cite{Jockers, Grimm-eff-F}, the adjoint charged matters 
of the first kind mix with Kahler moduli of the bulk Calabi--Yau, and 
the adjoints of the second kind with complex structure moduli (and
dilaton) of the bulk Calabi--Yau. 
Once a charged matter chiral multiplets on matter curves (D7--D7
intersection curves in Type IIB orientifold language) come into a
picture, however, such a separation may not be maintained; such 
charged matter fields are described by wavefunctions that do not 
vanish in either one of $\psi$ and $\chi$. They have character 
of both kinds. Yet, once we follow the Higgs cascade further 
down \cite{Higgs-cascade}
(while ignoring phenomenological applications) so that no non-Abelian 
symmetry is left unbroken and the F-theory Calabi--Yau 4-fold becomes 
smooth, the separation between the Kahler moduli Kahler potential and 
complex structure moduli Kahler potential will emerge, unless the 
bad approximation of the Kaluza--Klein {\it truncation} is to be blamed.
It is an interesting question how this happens.

  \section*{Acknowledgements}  

This work was supported in part 
by a Grant-in-Aid \#23540286 from the MEXT of Japan (TK), 
by Global COE Program ``the Physical Sciences Frontier'', MEXT, Japan (YT), 
by WPI Initiative and by a Grant-in-Aid for Scientific Research on
Innovative Areas 2303 from MEXT, Japan (TW).


%

\end{document}